\documentclass[superscriptaddress,aps,showpacs,nofootinbib,twocolumn]{revtex4}
\usepackage{graphicx}
\usepackage{amsmath,amssymb}

\def\CTP{{\it Commun. Theor. Phys.} }
\def\CQG{{\it Class. Quantum Gravity} }

\def\JHEP{{\it JHEP} }

\def\NAT{{\it Nature} }

\def\PL{{\it Phys. Lett.} }
\def\PR{{\it Phys. Rev.} }
\def\PRL{{\it Phys. Rev. Lett.} }
\def\PRTS{{\it Physics Reports} }

\def\prt{\partial}

\def\frac#1#2{{\textstyle{{#1}\over {#2}}}}

\def\lsim{\mathrel{\rlap{\lower4pt\hbox{\hskip1pt$\sim$}}
    \raise1pt\hbox{$<$}}}
\def\gsim{\mathrel{\rlap{\lower4pt\hbox{\hskip1pt$\sim$}}
    \raise1pt\hbox{$>$}}}
\def\sqr#1#2{{\vcenter{\vbox{\hrule height.#2pt
         \hbox{\vrule width.#2pt height#1pt \kern#1pt
         \vrule width.#2pt}
         \hrule height.#2pt}}}}

\def\beq{\begin{equation}}
\def\eeq{\end{equation}}
\def\beqa{\begin{eqnarray}} 
\def\eeqa{\end{eqnarray}}

\def\laq{\raise 0.4 ex \hbox{$<$}\kern -0.8 em\lower 0.62 ex\hbox{$\sim$}}
\def\gaq{\raise 0.4 ex \hbox{$>$}\kern -0.7 em\lower 0.62 ex\hbox{$\sim$}}

\begin{document}

\title{Noncommutative Gravitational Quantum Well}

\author{O. Bertolami}
\altaffiliation{Email address: orfeu@cosmos.ist.utl.pt}

\author{J. G. Rosa}
\altaffiliation{Email address: joaopedrotgr@sapo.pt}
 
\affiliation{ Departamento de F\'\i sica, Instituto Superior T\'ecnico \\
Avenida Rovisco Pais 1, 1049-001 Lisboa, Portugal}

\author{C. M. L. de Arag\~ao}
\altaffiliation{Email address: cristiane.aragao@ct.infn.it}

\author{P. Castorina}
\altaffiliation{Email address: paolo.castorina@ct.infn.it}

\author{D. Zappal\`a}
\altaffiliation{Email address: dario.zappala@ct.infn.it}

\affiliation{Department of Physics, University of Catania and INFN-Sezione di Catania, Citt\'a Universitaria, Via S. Sofia 64, Catania, Italy}

\vskip 0.5cm

\date{\today}

\begin{abstract}

We study noncommutative geometry at the Quantum Mechanics level by means of a model where noncommutativity of both configuration and momentum spaces is considered. We analyze how this model affects the problem of the two-dimensional gravitational quantum well and use the latest experimental results for the two lowest energy states of neutrons in the Earth's gravitational field to establish an upper bound on the fundamental momentum scale introduced by noncommutativity, namely $\sqrt{\eta}\lesssim1\ \mathrm{meV/c}$, a value that can be improved in the future by up to $3$ orders of magnitude. We show that the configuration space noncommutativity has, in leading order, no effect on the problem. We also analyze some features introduced by the model, specially a correction to the presently accepted value of Planck's constant to $1$ part in $10^{24}$. 

\end{abstract}

\pacs{32.80.Rm, 03.65.Ta, 11.10.Ef \hspace{2cm}Preprint DF/IST-2.2005} 

\maketitle
 

\section{Introduction}

The issue of noncommutative geometry has been extensively discussed in the recent literature. Although its study has a long standing story \cite{Snyder}, there has been a growing interest on this subject since the discovery in string theory that the low-energy effective theory of a D-brane in the background of a NS-NS B field lives on a noncommutative space \cite{Connes, Seiberg}. Furthermore, it has been suggested that the noncommutativity of our spacetime may arise as a quantum effect of gravity. Thus, noncommutative spaces provide a natural background for a possible regularization of quantum field theories \cite{Ho}. Furthermore, since brane fluctuations are described by gauge theories, the existence of noncommutative branes has motivated an extensive study of noncommutative gauge theories or, more generally, noncommutative field theories (NFCT's). These theories are based on the Weyl-Moyal correspondence, in which all products are replaced by the star-product in order to obtain their noncommutative action \cite{Szabo}. 

An important issue is that, in the case of space-time noncommutativity, the correspondent field theories are not unitary, which makes them less appealing. However, for light-like noncommutativity, there is still a well-defined quantum field theory (see \cite{Szabo, Bertolami_1} and references therein).

Another interesting feature of noncommutative geometry is its direct connection with the breaking of the Lorentz invariance \cite{Carroll}. The violation of the Lorentz symmetry arises directly from the commutator of the coordinates $x^{\mu}$ on the space-time manifold, which can be written as:
\beq \label{noncommutation_1}
[x^{\mu}, x^{\nu}]=i\theta^{\mu\nu}~.
\eeq

Under an observer Lorentz transformation, which includes boosts and/or rotations of the observer's inertial frame, Eq. (\ref{noncommutation_1}) transforms covariantly, as both the coordinates $x^{\mu}$ and the noncommutative parameter $\theta^{\mu\nu}$ transform as Lorentz tensors. However, under particle Lorentz transformations, which concern boosts and/or rotations of the matter fields, $\theta^{\mu\nu}$ behaves as the vacuum expectation value of some Lorentz tensor arising from the spontaneous symmetry breaking of the underlying fundamental theory. This means that this type of Lorentz transformation leaves $\theta^{\mu\nu}$ unaffected, while the commutator $[x^{\mu},x^{\nu}]$ transforms covariantly in the usual way. Thus, space-time noncommutativity directly implies the violation of the Lorentz invariance\footnote{Notice, however, that $\theta^{\mu\nu}$ can be regarded as a Lorentz tensor and Lorentz invariance may hold, at least at first non-trivial order in perturbation theory of the noncommutative parameter \cite{Bertolami_2}.}.

Although the effects of noncommutativity should presumably become significant at very high energy scales (close, for instance, to the string scale), it is fascinating to speculate whether there might be some low-energy effects of the fundamental quantum field theory. These effects might arise as a noncommutative version of Quantum Mechanics, NCQM, which lately has also been the focus of a lively discussion in the literature \cite{Ho, Nair, Zhang_1, Zhang_2, Demetrian, Gamboa}.

In this work, we study the phenomenology of a noncommutative model of Quantum Mechanics, in which noncommutativity of both configuration and momentum spaces is considered. It should be realized that the latter arises naturally as a consequence of the former, as the momenta are defined to be the partial derivatives of the action with respect to the noncommutative spatial coordinates \cite{Singh}. We analyze how the presence of both kinds of noncommutativity may affect the commutation relation between coordinates and momenta, leading to a redefinition of the Planck constant, $\hbar$.

We apply this NCQM model to the problem of a particle in the quantum well of the Earth's gravitational field to determine how noncommutativity affects its energy spectrum. We then use the experimental results recently obtained by Nesvizhevsky \emph{et al.} \cite{Nesvizhevsky_1, Nesvizhevsky_2} to place upper bounds on the noncommutative parameters of the NCQM model.

This paper is divided into five sections. In the next section, we describe the main features of the noncommutative model and determine how it modifies the Hamiltonian for the gravitational quantum well. In section III, we briefly describe the system's energy spectrum and wave functions in the commutative case, as well as the experiment of Refs. \cite{Nesvizhevsky_1, Nesvizhevsky_2}. In section IV, we derive the bounds on the noncommutative parameters that can be determined from the latest experimental results and analyze how these bounds can be improved in future experiments. Finally, in section V, we discuss the obtained results and some of the features of the model in comparison with other recently proposed noncommutative models.


\section{Noncommutative Quantum Mechanics}

In a model where both configuration and momentum space noncommutativity is considered, the coordinates and momenta must satisfy, in a 4-dimensional space, the following algebra:
\beqa \label{noncommutation_2}
\lbrack x^{\mu},x^{\nu}]&=&i\theta^{\mu\nu}\nonumber\\
\lbrack p^{\mu},p^{\nu}]&=&i\eta^{\mu\nu}\nonumber\\
\lbrack x^{\mu},p^{\nu}]&=&i\hbar\delta^{\mu\nu}~,
\eeqa
where the parameters $\theta^{\mu\nu}$ and $\eta^{\mu\nu}$ are antisymmetric. This algebra is consistent with usual Quantum Mechanics through the last commutation relation in Eq. (\ref{noncommutation_2}).

In this work, we shall not consider time-like noncommutative relations, i.e., we take $\theta^{0i}=0$ and $\eta^{0i}=0$, as otherwise the underlying quantum field theory is not unitary, as previously referred to. Since the system in which we study the effects of noncommutativity is two-dimensional, we limit our analysis to the $xy$ plane, where the noncommutative algebra can be written as:
\beqa \label{noncommutation_3}
\lbrack x,y]&=&i\theta\nonumber\\
\lbrack p_x,p_y]&=&i\eta\nonumber\\
\lbrack x_i,p_j]&=&i\hbar\delta_{ij}\qquad i=1,2~,
\eeqa
where, in the last commutation relation, we identified $x_1\equiv x$, $x_2\equiv y$, $p_1\equiv p_x$ and $p_2\equiv p_y$. 

The parameters $\theta$ and $\eta$ correspond to the components $\theta^{12}$ and $\eta^{12}$ of the noncommutative parameters in Eq. (\ref{noncommutation_2}). As such, they correspond to vacuum expectation values of the components of some Lorentz tensors, making them invariant under particle Lorentz transformations. However, they are not invariant under a boost or rotation of the observer's inertial frame. This implies that, if the observer's inertial frame exhibits a motion which depends on the space-time coordinates, then $\theta$ and $\eta$ will be also space-time dependent. For now, we consider that both parameters are constant throughout all space-time. In section V, we will show that this is, in fact, a good approximation at least within the framework of the experiment analyzed in this work.

The commutation relations Eq. (\ref{noncommutation_3}) lead to the following uncertainty relations:
\beqa \label{uncertainty_1}
\Delta x\Delta y&\geq&{\theta\over 2}\nonumber\\
\Delta p_x\Delta p_y&\geq&{\eta\over 2}\nonumber\\
\Delta x\Delta p_x&\geq&{\hbar\over 2}\nonumber\\
\Delta y\Delta p_y&\geq&{\hbar\over 2}~.
\eeqa

These relations arise from the general uncertainty principle which states that, for two hermitian operators $A$ and $B$, $\Delta A\Delta B\geq\langle i[A,B]\rangle/2$ \cite{Gasiorowicz}. The first two uncertainty relations show that measurements of positions and momenta in both directions $x$ and $y$ are not independent. Taking into account the fact that $\theta$ and $\eta$ have dimensions of $(length)^2$ and $(momentum)^2$, respectively, then $\sqrt{\theta}$ and $\sqrt{\eta}$ define fundamental scales of length and momentum which characterize the minimum uncertainties possible to achieve in measuring these quantities. One expects these fundamental scales to be related to the scale of the underlying field theory (possibly the string scale) and, thus, to appear as small corrections at the low-energy level of Quantum Mechanics.

One possible way of implementing algebra Eq. (\ref{noncommutation_3}) is to construct the noncommutative variables $\{x,y,p_x,p_y\}$ from the commutative variables $\{x',y',p_x',p_y'\}$ by means of linear transformations. Given the canonical Heisenberg commutation relations,
\beqa \label{heisenberg}
\lbrack x',y']&=&[p_x',p_y']=0\nonumber\\
\lbrack x'_i,p'_j]&=&i\hbar\delta_{ij}\qquad i=1,2~,
\eeqa
one can easily verify that the commutation relations Eq. (\ref{noncommutation_3}) can be obtained through the linear transformations:
\beqa \label{linear_1}
x&=&x'-{\theta\over\hbar}p_y'\nonumber\\
y&=&y'\nonumber\\
p_x&=&p_x'\nonumber\\
p_y&=&p_y'-{\eta\over\hbar}x'~,
\eeqa
or through the linear transformations:
\beqa \label{linear_2}
x&=&x'\nonumber\\
y&=&y'+{\theta\over\hbar}p_x'\nonumber\\
p_x&=&p_x'+{\eta\over\hbar}y'\nonumber\\
p_y&=&p_y'~.
\eeqa

In the case of a two-dimensional system for which the Hamiltonian remains invariant under a rotation of $\pi/2$, both types of linear transformations Eqs. (\ref{linear_1}) and (\ref{linear_2}) are equivalent, and the implementation of noncommutativity can be achieved through either of them. However, for a system that does not exhibit such a symmetry, as the one we will study in this work, the two types of linear transformations will lead to different results. This means that, in general, the choice of the linear transformations through which to implement noncommutativity is ambiguous. This ambiguity turns the noncommutative model Eq. (\ref{noncommutation_3}) into an ill-defined problem that cannot correspond to any physical reality.

The simplest way to solve this problem would be to combine the two types of linear transformations into a single one that simultaneously modifies all coordinates and momenta, and not just $x$ and $p_y$ or $y$ and $p_x$ as in Eqs. (\ref{linear_1}) and (\ref{linear_2}). We find that through the linear transformations:
\beqa \label{linear_3}
x&=&x'-{\theta\over2\hbar}p_y'\nonumber\\
y&=&y'+{\theta\over2\hbar}p_x'\nonumber\\
p_x&=&p_x'+{\eta\over2\hbar}y'\nonumber\\
p_y&=&p_y'-{\eta\over2\hbar}x'~,
\eeqa
the two first commutation relations in Eq. (\ref{noncommutation_3}) are recovered. However, the last one is changed to:
\beq \label{noncommutation_4}
[x_i,p_j]=i\hbar\bigg(1+{\theta\eta\over4\hbar^2}\bigg)\delta_{ij}\qquad i=1,2~.
\eeq

Comparing Eqs. (\ref{noncommutation_4}) and (\ref{noncommutation_3}), we find that the linear transformations Eq. (\ref{linear_3}) lead to the appearance of an \emph{effective Planck constant} which depends on the noncommutative parameters $\theta$ and $\eta$, and is given by:
\beq \label{Planck_constant_1}
\hbar_{eff}=\hbar(1+\xi)~,
\eeq
where $\xi\equiv \theta\eta/4\hbar^2$. For sure, this setup is consistent with the usual commutative space-time Quantum Mechanics only if $\xi\ll 1$. This is expected to be the case as $\xi$ is of second order on the noncommutative parameters $\theta$ and $\eta$. The results of the experiment by Nesvizhevsky \emph{et al.} will allow us to estimate an upper bound for the value of $\xi$ and to evaluate the consistency of the noncommutative model. From now on, we can assume that the NCQM problem in hand is well-defined. We mention that in Ref. \cite{Singh}, an effective Planck constant also appears as a consequence of both space and momentum noncommutativity, even though within a somewhat different framework.

For completeness, we present the generalization of Eq. (\ref{linear_3}) to the 4-dimensional space:
\beqa \label{linear_4}
x^{\mu}&=&x'^{\mu}-{\theta^{\mu}_{\ \nu}\over2\hbar}p'^{\nu}\nonumber\\
p^{\mu}&=&p'^{\mu}+{\eta^{\mu}_{\ \nu}\over2\hbar}x'^{\nu}~.
\eeqa

This leads to the following commutation relations:
\beqa \label{noncommutation_5}
\lbrack x^{\mu},x^{\nu}]&=&i\theta^{\mu\nu}\nonumber\\
\lbrack p^{\mu},p^{\nu}]&=&i\eta^{\mu\nu}\nonumber\\
\lbrack x^{\mu},p^{\nu}]&=&i\hbar\bigg(\delta^{\mu\nu}+{\theta^{\mu\alpha}\eta^{\nu}_{\ \alpha}\over4\hbar^2}\bigg)~.
\eeqa

Notice that a modification like this was discussed in Ref. \cite{Djemai}. Hence, we see that in the 4-dimensional case not only there is an \emph{effective Planck constant} given by
\beq \label{Planck_constant_2}
\hbar_{eff}=\hbar\bigg(1+{Tr[\theta\eta]\over4\hbar^2}\bigg)~,
\eeq
but also that $[x^{\mu},p^{\nu}]$ is no longer diagonal, with the off-diagonal elements being proportional to products between components of $\theta^{\mu\nu}$ and $\eta^{\mu\nu}$. One should notice that neither the effective Planck constant nor the off-diagonal elements appear in the case where only noncommutativity between coordinates is considered.

Turning back to the two-dimensional case, it is clear that we need the inverse transformations of Eq. (\ref{linear_3}) in order to convert a commutative Hamiltonian into a noncommutative one. This is given by:
\beqa \label{linear_inverse}
x'&=&C\bigg(x+{\theta\over2\hbar}p_y\bigg)\nonumber\\
y'&=&C\bigg(y-{\theta\over2\hbar}p_x\bigg)\nonumber\\
p_x'&=&C\bigg(p_x-{\eta\over2\hbar}y\bigg)\nonumber\\
p_y'&=&C\bigg(p_y+{\eta\over2\hbar}x\bigg)~,
\eeqa
where we have set $C\equiv(1-\xi)^{-1}$.

Consider now a system of a particle of mass $m$ moving in the $xy$ plane, subject to the Earth's gravitational field, $\mathbf{g}=-g\mathbf{e_x}$, where $g\simeq9.81\ \mathrm{ms^{-2}}$ is assumed to be constant near its surface. This corresponds to the experimental setup of Refs. \cite{Nesvizhevsky_1, Nesvizhevsky_2}. The system's Hamiltonian is, in the commutative case, given by:
\beq \label{Hamiltonian_1}
H'={p_x'^2\over2m}+{p_y'^2\over2m}+mgx'~.
\eeq

The corresponding noncommutative Hamiltonian can be straightforwardly obtained using the inverse transformations Eq. (\ref{linear_inverse}) to replace the commutative variables by the noncommutative ones:
\beqa \label{Hamiltonian_2}
H&=&{C^2\over2m}\big(p_x-{\eta\over2\hbar}y\big)^2+{C^2\over2m}\big(p_y+{\eta\over2\hbar}x\big)^2+\nonumber\\
&+&mgC\big(x+{\theta\over2\hbar}p_y\big)=\nonumber\\
&=&{C^2\over2m}p_x^2+{C^2\over2m}p_y^2+mgC{\theta\over2\hbar}p_y+\nonumber\\
&+&{C^2\over2m}{\eta\over\hbar}(xp_y-yp_x)+{C^2\over8m\hbar^2}\eta^2(x^2+y^2)+\nonumber\\
&+&mgCx~.
\eeqa

One should notice that
\beq \label{aux_1}
{C^2\over2m}p_y^2+mgC{\theta\over2\hbar}p_y={1\over2m}\bigg(Cp_y+{m^2g\theta\over2\hbar}\bigg)^2-{m^3g^2\theta^2\over8\hbar^2}~,
\eeq
where the last term is an additive constant that can be removed from the Hamiltonian.

By defining
\beq \label{momenta}
\bar{p_x}\equiv Cp_x\qquad\bar{p_y}\equiv Cp_y+{m^2g\theta\over2\hbar}~,
\eeq 
the noncommutative Hamiltonian can be written as:
\beqa \label{Hamiltonian_3}
H&=&{\bar{p_x}^2\over2m}+{\bar{p_y}^2\over2m}+{C\eta\over2m\hbar}(x\bar{p_y}-y\bar{p_x})+\nonumber\\
&+&{C^2\over8m\hbar^2}\eta^2(x^2+y^2)+mgCx-mgC{\theta\eta\over4\hbar^2}x~.
\eeqa

Notice that the last two terms correspond to the commutative gravitational potential:
\beq \label{aux_2}
mgCx-mgC{\theta\eta\over4\hbar^2}x=mgC(1-\xi)x=mgx~.
\eeq

Thus, the noncommutative Hamiltonian is given by:
\beqa \label{Hamiltonian_4}
H&=&{\bar{p_x}^2\over2m}+{\bar{p_y}^2\over2m}+mgx+\nonumber\\
&+&{C\eta\over2m\hbar}(x\bar{p_y}-y\bar{p_x})+{C^2\over8m\hbar^2}\eta^2(x^2+y^2)~.
\eeqa

The similarity between the first three terms in the commutative and noncommutative Hamiltonians is evident, the only difference lying in the redefined momenta $\bar{p_x}$ and $\bar{p_y}$. The constant term in the definition of $\bar{p_y}$, Eq. (\ref{momenta}), proportional to $\theta$, has no physical meaning, as it only produces a translation of all the eigenvalues of $Cp_y$ by the same amount. It does not introduce any modifications whatsoever on the commutation relations of this operator either. Hence, the only physical difference between the operators $p_y$ and $\bar{p_y}$ resides on the factor $C$, as happens with $p_x$ and $\bar{p_x}$. Thus, the kinetic terms in the noncommutative Hamiltonian differ from the commutative ones by a factor $C^2$.

Before we go further into the study of this noncommutative Hamiltonian, we shall briefly describe in the next section the solutions to this problem in the commutative case.


\section{The Gravitational Quantum Well}

We consider now the experiment described in Refs. \cite{Nesvizhevsky_1, Nesvizhevsky_2}. In the case where a horizontal mirror is placed at $x=0$, a quantum well is formed by the mirror and the constant gravitational field. This system is known as the \emph{gravitational quantum well}. The solutions to the eigenvalue equation in the commutative case, $H'\psi_n=E_n\psi_n$, are well known \cite{Landau}. The system's wave function can be separated into two parts, corresponding to each of the commutative coordinates $x'$ and $y'$. The eigenfunctions corresponding to $x'$ can be expressed in terms of the Airy function $\phi(z)$\footnote{This function corresponds to the Airy function $Ai(z)$, which is normalizable. This does not happen with the Airy function $Bi(z)$, which is also a solution to the problem.},
\beq \label{Airy_1}
\psi_n(x')=A_n\phi(z)~,
\eeq
with eigenvalues determined by the roots of the Airy function, $\alpha_n$, with $n=1,2\ldots$,
\beq \label{Airy_2}
E_n=-\bigg({mg^2\hbar^2\over2}\bigg)^{1/3}\alpha_n~.
\eeq

The variable $z$ is related to the height $x'$ by means of the following linear relation:
\beq \label{Airy_3}
z=\bigg({2m^2g\over\hbar^2}\bigg)^{1/3}\bigg(x'-{E_n\over mg}\bigg)~.
\eeq

The normalization factor for the $n$-th eigenstate is given by:
\beq \label{Airy_4}
A_n=\Bigg[\bigg({\hbar^2\over 2m^2g}\bigg)^{1\over3}\int_{\alpha_n}^{+\infty}dz\phi^2(z)\Bigg]^{-{1\over2}}~.
\eeq

In what concerns the part of the particle's wave function in the horizontal direction $y'$, one can easily deduce that it corresponds to a group of plane waves with a continuous energy spectrum, as the particle is free in the direction transverse to the gravitational potential:
\beq \label{Airy_5}
\psi(y')=\int_{-\infty}^{+\infty}g(k)e^{iky'}dk~,
\eeq
where the function $g(k)$ determines the group's shape in phase space.

As a consequence of the particle's discrete energy spectrum in the gravitational field's direction, the probability of finding the particle at a certain height will be maximum for the classical turning point $x_n=E_n/mg$ for each quantum state. This height corresponds to the classically allowed height for a particle with energy $E_n$\footnote{The extent to which the Equivalence Principle can be said to hold in this experiment is discussed in Ref. \cite{Bertolami_3}.}. As soon as the particle's height exceeds this value, the probability of finding it starts to decay exponentially \cite{Nesvizhevsky_1, Nesvizhevsky_2}.

This property allowed Nesvizhevsky \emph{et al.} to identify the quantum states of neutrons in the quantum well formed by the Earth's gravitational field and a horizontal mirror. By placing a scatterer/absorber above the horizontal mirror, they were able to measure the neutron transmission through the narrow slit between them. If the height of the scatterer/absorber is much higher than the classical turning point for a given quantum state, the neutrons pass through the slit without significant losses. As the slit size decreases, the probability of neutron loss will increase until the slit height approaches $x_n$ and the slit stops being transparent to neutrons. A more detailed description of the experimental apparatus and procedure, as well as the report of the first identification of the lowest quantum state, can be found in Refs. \cite{Nesvizhevsky_1}.

Clearly, the choice of neutrons for the experiment is to avoid that electromagnetic effects overlap the effect of the Earth's gravitational field on the energy spectrum. Neutron's long lifetime ($\tau\simeq885.7\:\mathrm{s}$) \cite{PDG} and mass ($m_n\simeq939.57\:\mathrm{MeV/c^2}$) also bring some advantages to the performance of the experiment.

Recently, Nesvizhevsky \emph{et al.} \cite{Nesvizhevsky_2} were able to determine the values of the classical heights for the first two quantum states, obtaining the following results:
\beqa \label{Nesvizhevsky_1}
x_1^{exp}&=&12.2\pm1.8(syst.)\pm0.7(stat.)\ (\mathrm{\mu m})~,\nonumber\\
x_2^{exp}&=&21.6\pm2.2(syst.)\pm0.7(stat.)\ (\mathrm{\mu m})~.
\eeqa

The corresponding theoretical values can be determined from Eq. (\ref{Airy_2}) for $\alpha_1=-2.338$ and $\alpha_2=-4.088$, yielding $x_1=13.7\:\mathrm{\mu m}$ and $x_2=24.0\:\mathrm{\mu m}$, corresponding to the energy eigenvalues $E_1=1.407\:\mathrm{peV}$ and $E_2=2.461\:\mathrm{peV}$. These values are contained in the error bars and allow for maximum absolute shifts of the energy levels with respect to the predicted values:
\beqa \label{Nesvizhesky_2}
\Delta E_1^{exp}&=&6.55\times10^{-32}\ \mathrm{J}=0.41\ \mathrm{peV}~,\nonumber\\
\Delta E_2^{exp}&=&8.68\times10^{-32}\ \mathrm{J}=0.54\ \mathrm{peV}~.
\eeqa

In this experiment, neutrons exhibited a mean horizontal velocity of $\langle v_y\rangle\simeq6.5\:\mathrm{ms^{-1}}$.


\section{Bounds on the NCQM parameters}

Consider now the noncommutative case, for which the Hamiltonian is given by Eq. (\ref{Hamiltonian_4}). To first order in the noncommutative parameters $\theta$ and $\eta$, the Hamiltonian is approximately given by:
\beqa \label{Hamiltonian_5}
H&=&{p_x^2\over2m}+{p_y^2\over2m}+mgx+{\eta\over2m\hbar}(xp_y-yp_x)=\nonumber\\
&=&H'+{\eta\over2m\hbar}(xp_y-yp_x)~.
\eeqa

Notice that $C=1+\xi+O((\theta\eta)^2)$, so that, to first order on the noncommutative parameters, $\bar{p_x}$ and $\bar{p_y}$ are equal to $p_x$ and $p_y$, respectively. Thus, we conclude that the noncommutative Hamiltonian differs from the commutative one by a term proportional to $\eta$ at this order of approximation. It then follows that the configuration space noncommutativity does not influence the gravitational quantum well energy spectrum to leading order.

As previously referred to, $\eta$ must be a small correction at the low-energy level, and so we can treat the new term as a perturbation in the commutative Hamiltonian. The shift caused by this term on the system's energy levels is given by the expectation value of the perturbation on the system's wave function. We first point out that, as the Airy function is real, $\psi_n(x)=\psi_n^{*}(x)$, therefore
\beqa \label{aux_3}
\langle p_x \rangle_n&=&\int_0^{+\infty}dx\:\psi_n^{*}\bigg(-i\hbar{\prt\over\prt x}\psi_n\bigg)=\nonumber\\
&=&-i\hbar\Big([\psi_n^{*}\psi_n]_{0}^{+\infty}-\int_0^{+\infty}dx\:{\prt\psi_n^{*}\over\prt x}\psi_n\Big)=\nonumber\\
&=&i\hbar\int_0^{+\infty}dx\:\psi_n{\prt\psi_n\over\prt x}=\nonumber\\
&=&-\int_0^{+\infty}dx\:\psi_n^{*}\bigg(-i\hbar{\prt\over\prt x}\psi_n\bigg)=\nonumber\\
&=&-\langle p_x\rangle_n=0~,
\eeqa
where we have used the fact that $\psi_n(x=0)=0$, due do the presence of the horizontal mirror, and the normalizability of the wave function. Thus, the term proportional to $p_x$ in Eq. (\ref{Hamiltonian_5}) will not produce any shift on the system's energy levels, whatever the expectation value of $y$\footnote{As long as it is a finite value, which is expected for a localized group of plane waves.}. Thus, the leading order perturbative potential due to noncommutativity is then given by:
\beq \label{perturbative_1}
V_1={\eta\over2m\hbar}xp_y~.
\eeq

This is clearly analogous to a potential describing the effect of a magnetic field $\mathbf{B}=B\mathbf{e_z}$, where $z$ is the direction perpendicular to the plane, on a particle of charge $q$, with the identification $qB=\eta/2\hbar$. We point out that this is simply a formal analogy with no physical meaning, as particles in the gravitational quantum well may be neutral, as is the case of the neutrons used in the experiment by Nesvizhevsky \emph{et al.}.

The leading order energy correction to the $n$-th quantum state is given by the expectation value of potential Eq. (\ref{perturbative_1}), which can be written as:
\beqa \label{perturbative_2}
\Delta E_n^{(1)}&=&{\eta k\over2m}\int_0^{+\infty}dx\psi_n^*(x)x\psi_n(x)=\nonumber\\
&=&{\eta k\over2m}\Bigg[\bigg({2m^2g\over\hbar^2}\bigg)^{-\frac{2}{3}}A_n^2I_n+{E_n\over mg}\Bigg]~,
\eeqa
where the integral $I_n$ is defined as:
\beq \label{aux_4}
I_n\equiv\int_{\alpha_n}^{+\infty}dz\phi(z)z\phi(z)~,
\eeq
and $k=\langle p_y\rangle/\hbar=m\langle v_y\rangle/\hbar=1.03\times10^8\:\mathrm{m^{-1}}$ for the experiment by Nesvizhevsky \emph{et al.} \cite{Nesvizhevsky_2}. The values of the normalization factor $A_n$ and of the integral $I_n$ were numerically determined for the first two energy levels, that is:
\beqa \label{numerical_1}
&A_1= 588.109\ ,\qquad & I_1=-0.383213~,\nonumber\\
&A_2= 513.489\ ,\qquad & I_2=-0.878893~.
\eeqa

With these values, the leading order corrections to the energy levels are given by:
\beqa \label{energy_corrections}
\Delta E_1^{(1)}=2.83\times10^{29}\eta\qquad\mathrm{(J)~,}\nonumber\\
\Delta E_2^{(1)}=4.94\times10^{29}\eta\qquad\mathrm{(J)~.}
\eeqa

By requiring these corrections to be smaller or of the order of the maximum absolute energy shifts allowed by the experiment, we obtain the following upper bounds for the value of $\eta$:
\beqa \label{eta_bounds_1}
|\eta|&\lesssim& 2.32\times 10^{-61}\ \mathrm{kg^2m^2s^{-2}}\qquad(n=1)~,\nonumber\\
|\eta|&\lesssim& 1.76\times 10^{-61}\ \mathrm{kg^2m^2s^{-2}}\qquad(n=2)~.
\eeqa

These values correspond to the following upper bounds on the fundamental momentum scale:
\beqa \label{eta_bounds_2}
|\sqrt{\eta}|&\lesssim& 4.82\times10^{-31}\:\mathrm{kgms^{-1}}\nonumber\\
&\lesssim&0.90\:\mathrm{meV/c}\qquad\qquad(n=1)~,\nonumber\\
|\sqrt{\eta}|&\lesssim& 4.20\times10^{-31}\:\mathrm{kgms^{-1}}\nonumber\\
&\lesssim&0.79\:\mathrm{meV/c}\qquad\qquad(n=2)~.
\eeqa

We determine now the energy correction of second order on the noncommutative parameters. The second order perturbative potential is given by:
\beq \label{perturbative_3}
V_2={\theta\eta\over2\hbar^2}{p_x^2\over2m}+{\theta\eta\over2\hbar^2}{p_y^2\over2m}+{\eta^2\over8m\hbar^2}(x^2+y^2)~.
\eeq

The terms proportional to $p_y^2$ and $y^2$ do not affect the particle's energy spectrum in the direction of the gravitational field and, thus, do not produce any shifts on the discrete energy levels. Hence, the second order perturbative potential reduces to:
\beq \label{perturbative_4}
V_2={\theta\eta\over2\hbar^2}{p_x^2\over2m}+{\eta^2\over8m\hbar^2}x^2~.
\eeq

There are, thus, two second order terms that can modify the particle's energy spectrum. The first one is proportional to the particle's kinetic energy in the direction of the gravitational field; the second one is formally identical to an harmonic oscillator with frequency $\omega=|\eta|/2m\hbar$\footnote{Once again, we emphasize the fact that this is just a formal analogy with no physical meaning.}. The energy correction due to the first term on the $n$-th quantum state is given by:
\beqa \label{perturbative_5}
\Delta E_n^{(2a)}&=&-{\theta\eta\over4m}\int_0^{+\infty} dx\:  \psi_n^{*}(x){\prt^2\psi_n\over\prt x^2}(x)\nonumber\\
&=&-{\theta\eta\over4m}A_n^2\bigg({2m^2g\over\hbar^2}\bigg)^{1\over3}J_n~,
\eeqa
where the integral $J_n$ is defined as:
\beq \label{aux_5}
J_n\equiv\int_{\alpha_n}^{+\infty}dz\phi(z){d^2\phi\over dz^2}(z)~.
\eeq

We have determined the value of this integral numerically for the first two quantum states, obtaining the following results:
\beq \label{numerical_2}
J_1=-0.383213\qquad,\qquad J_2=-0.878893~.
\eeq

In order to set an upper bound on the value of this correction, we need not only the upper bounds obtained for 
$\eta$ but also an upper bound for the value of $\theta$, which cannot be estimated by the graviatational 
quantum well experiment. Then, one can resort to 
the bound on the value of the coordinates commutator,
derived in a different context \cite{Carroll}, $\theta\simeq 4\times10^{-40} m^2$ (which correspond to 
$\theta\simeq(10~TeV)^{-2}$ for $\hbar=c=1$)\footnote{Another bound can be found, for instance, 
in Ref. \cite{Bertolami_1}.}, 
or, otherwise, one can assume a much more conservative point of view 
and argue that the fundamental length scale introduced by noncommutativity in our specific case should be at least smaller than 
the minimum scale compatible with the quantum mechanical approach to the gravitational quantum well problem. 
This scale is given by the average neutron size of $\sim 1\:\mathrm{fm}$, below which the neutron's internal 
structure becomes significant. With this latter  hypothesis, one can get an upper bound on 
$\theta$ of $10^{-30}\:\mathrm{m^2}$ and, consequently, the following upper bounds on the 
contribution of Eq. (\ref{perturbative_5}) to the energy correction:
\beqa \label{perturbative_6}
\Delta E_1^{(2a)}&\lesssim&7.83\times10^{-55}\:\mathrm{(J)}~,\nonumber\\
\Delta E_2^{(2a)}&\lesssim&1.04\times10^{-54}\:\mathrm{(J)}~.
\eeqa

As for the contribution of the second term, it is given by:
\beqa \label{perturbative_7}
\Delta E_n^{(2b)}&=&{\eta^2\over8m\hbar^2}\int_0^{+\infty}dx\:\psi_n^{*}(x)x^2\psi_n(x)=\nonumber\\
&=&{\eta^2\over8m\hbar^2}\Bigg[\bigg({2m^2g\over\hbar^2}\bigg)^{-1}A_n^2L_n+\nonumber\\
&+&\bigg({2m^2g\over\hbar^2}\bigg)^{-\frac{2}{3}}{2E_n\over mg}A_n^2I_n+\bigg({E_n\over mg}\bigg)^2\Bigg]~,
\eeqa
where the integral $L_n$ is defined by:
\beq \label{aux_6}
L_n\equiv\int_{\alpha_n}^{+\infty}dz\:\phi(z)z^2\phi(z)~,
\eeq
whose values were numerically determined for the first two energy levels:
\beqa \label{numerical_3}
L_1=0.537596\qquad,\qquad L_2=2.15572~.
\eeqa

Hence, 
\beqa \label{perturbative_8}
\Delta E_1^{(2b)}&\lesssim&3.64\times10^{-38}\:\mathrm{(J)}~,\nonumber\\
\Delta E_2^{(2b)}&\lesssim&6.39\times10^{-38}\:\mathrm{(J)}~.
\eeqa

Thus, we see that the contribution of the first set of second order terms is negligible in comparison with the contribution of the second term, which is itself $7\ (6)$ orders of magnitude smaller than the respective first order correction for $n=1$ ($n=2$). The perturbative approach is, thus, valid using the upper bounds obtained for $\eta$ for both quantum states. Clearly, had we used the bound on $\theta$ derived in \cite{Carroll}, the energy corrections would have been about ten orders of magnitude 
smaller, i.e. even more negligible than with the conservative bound used  for $\theta$.

Hence, the results of Nesvizhevsky \emph{et al.} constrain the fundamental momentum scale to be below the meV/c scale. However, one could expect the fundamental scale to be smaller than this. An increase in the precision of the experiment, which presently allows relatively large error bars, may lead to more stringent bounds on the value of $\sqrt{\eta}$  if the results are still consistent with the theoretical predictions. One should take into account, however, that the experimental energy resolution is bounded by the Uncertainty Principle due to the finite lifetime of the neutron \cite{Nesvizhevsky_1}. The maximum energy resolution that can be achieved corresponds to a minimum absolute energy uncertainty given by:
\beqa \label{uncertainty_2}
\Delta E^{min}&\sim&{\hbar\over\tau}\simeq1.2\times10^{-37}\ \mathrm{J}\simeq7.4\times10^{-19}\ \mathrm{eV}\nonumber\\
&\sim&10^{-18}\ \mathrm{eV}~.
\eeqa

If the theoretical predictions are confirmed by the experiment with this precision, then one should be able to place the following upper bounds on the value of $\eta$:
\beqa \label{eta_bounds_3}
|\eta|&\lesssim& 5.22\times 10^{-67}\ \mathrm{kg^2m^2s^{-2}}\qquad(n=1)~,\nonumber\\
|\eta|&\lesssim& 2.40\times 10^{-67}\ \mathrm{kg^2m^2s^{-2}}\qquad(n=2)~,
\eeqa
which correspond to the following upper bounds on the value of the fundamental momentum scale:
\beqa \label{eta_bounds_4}
|\sqrt{\eta}|&\lesssim& 7.22\times10^{-34}\:\mathrm{kgms^{-1}}\nonumber\\
&\lesssim&1.35\:\mathrm{\mu eV/c}\qquad\qquad(n=1)~,\\
|\sqrt{\eta}|&\lesssim& 4.90\times10^{-34}\:\mathrm{kgms^{-1}}\nonumber\\
&\lesssim&0.92\:\mathrm{\mu eV/c}\qquad\qquad(n=2)~.
\eeqa

These are the most stringent bounds that may be obtained within the framework of the quantum gravitational well.


\section{Discussion and outlook}

The results obtained in the previous section allow us to evaluate the consistency of the noncommutative model, i.e., whether the $\xi\ll 1$ hypothesis is consistent in the context of the results of the experiment by Nesvizhevsky \emph{et al.}. With the upper bound on $\theta$ of $10^{-30}\:\mathrm{m^2}$ and the upper bounds on $\eta$, Eqs. (\ref{eta_bounds_1}), that were obtained from the experimental error bars of the Nesvizhevsky \emph{et al.} experiment, we obtain the following bounds on the value of $\xi$:
\beqa \label{xi_1}
|\xi|&\lesssim& 5.2\times10^{-24}\qquad(n=1)~,\\
|\xi|&\lesssim& 4.0\times10^{-24}\qquad(n=2)~.
\eeqa

Hence, we can conclude that the modifications introduced by noncommutativity on the value of $\hbar$ are at least about $24$ orders of magnitude smaller than its value, which is known with a precision of about $10^{-9}$ \cite{PDG}. If we consider the most stringent bounds on $\eta$ obtained via Heisenberg's Uncertainty Principle, Eqs. (\ref{eta_bounds_3}), then:
\beqa \label{xi_2}
|\xi|&\lesssim& 1.2\times10^{-29}\qquad(n=1)~,\\
|\xi|&\lesssim& 5.4\times10^{-30}\qquad(n=2)~.
\eeqa

All the bounds on $\xi$ become about 10 orders of magnitude smaller if we consider the upper 
bound $\theta \simeq(10~TeV)^{-2}$ of Ref. \cite{Carroll}. Notice that these bounds also provide an 
estimate of the 4-dimensional correction to the Planck constant, which is about the same order of 
magnitude as the two-dimensional one\footnote{Assuming all components of $\theta^{\mu\nu}$ are of 
the same order of magnitude, the same happening to the components of $\eta^{\mu\nu}$.}. Hence, the 
noncommutative model we have considered in this work is consistent with all experimental results of ordinary Quantum Mechanics.

As previously mentioned, there is the possibility that the noncommutative parameters could exhibit a dependence on the space-time coordinates (see, e.g., Refs. \cite{Bertolami_1,Bertolami_2}). This occurs whether the laboratory frame has a space-time dependent motion. As proposed in Ref. \cite{Kamoshita}, it is likely that both $\theta^{\mu\nu}$ and $\eta^{\mu\nu}$ have fixed values in the Cosmic Microwave Background Radiation frame, which may be considered as approximately fixed to the celestial sphere. Therefore, physical measurements must take into account the effect of the Earth's rotation about its axis, which will yield a time dependence of the noncommutative parameters. 

In the case of the experiment by Nesvizhevsky \emph{et al.}, this would, however, have a negligible effect within the time scale of the experiment, which is limited by the neutron's finite lifetime, a mere $1\%$ of the Earth's rotation period. Thus, assuming that the noncommutative parameters are constant throughout the experiment seems to be a quite good approximation. Nevertheless, for sure, our results refer to the maximum effect that the parameter $\eta$ may have had on the system during the time frame of the experiment.

An important related issue is whether the effective Planck constant exhibits such a time dependence. From Eq. (\ref{Planck_constant_1}), we see that, if $\theta$ and $\eta$ are time-dependent, then $\hbar_{eff}$ should also depend on the time coordinate. However, one must not forget that Eq. (\ref{Planck_constant_1}) refers to the correction to $\hbar$ from a two-dimensional model. The true 4-dimensional effective Planck constant is given by Eq. (\ref{Planck_constant_2}). In this case, the correction behaves as a Lorentz scalar  both in  particle and observer Lorentz transformations, meaning that its value is the same in all inertial frames. Therefore, it can be concluded that the time dependence of $\xi$ will be canceled by the contraction and by the time dependence of components of $\theta^{\mu\nu}$ and $\eta^{\mu\nu}$. Hence, $\hbar_{eff}$ is a true constant and constitutes a fundamental noncommutative parameter that is valid for all inertial frames.

Recently, a relation between the noncommutative parameters has been proposed based on the assumption that the Bose-Einstein statistics is unaffected in noncommutative quantum mechanics \cite{Zhang_1}. To obtain this relation, one constructs the usual creation and annihilation operators for a two-dimensional isotropic harmonic oscillator of mass $m$ and frequency $\omega$ in terms of the noncommutative coordinates and momenta in the plane:
\beqa \label{creation_anihilation}
a_i&=&\sqrt{{m\omega\over2}}\bigg(x_i+{i\over m\omega}p_i\bigg)~,\\
a_i^{\dagger}&=&\sqrt{{m\omega\over2}}\bigg(x_i-{i\over m\omega}p_i\bigg)\qquad i=1,2~.
\eeqa

In order to retain the Bose-Einstein statistics, generated by creation and annihilation operators in commutative phase space, one should require that $[a_i,a_i^{\dagger}]=0$ in noncommutative phase space. This condition leads to the relation:
\beq \label{theta_eta_BE}
\eta=(m\omega)^2\theta~.
\eeq

Eq. (\ref{theta_eta_BE}) yields a direct proportionality between the two noncommutative parameters on the plane. This relation also exhibits a dependence on the parameters $m$ and $\omega$ of the two-dimensional isotropic harmonic oscillator. In many systems, the potential can be modelled by an harmonic oscillator through an expansion about its minimum. This is not, however, a general case. For instance, the linear potential appearing in the gravitational quantum well analyzed in this work, $mgx$, has no minimum, and its second derivative, which would correspond to its approximate harmonic oscillator frequency $\omega$, has a null value for all $x$. Thus, in our opinion, Eq. (\ref{theta_eta_BE}) has a limited applicability and, for sure, cannot be applied to the gravitational quantum well.

The dependence on the features of a particular system could also appear in the model considered in this work without great harm, as far as the invariance of $\hbar_{eff}$ under Lorentz transformations is not spoiled. This dependence could arise from the low-energy limit of the underlying quantum field theory. However, a relation between the noncommutative parameters should be valid for all systems, which is not the case of the underlying assumption behind Eq. (\ref{theta_eta_BE}).

Finally, we would like to point out that the bosonic creation and annihilation operators are not really constructed from the two-dimensional isotropic harmonic oscillator ones. Both types of operators are independently built and satisfy the same commutation relations in commutative phase space. However, this relation cannot be extended to the noncommutative case, where there is just an accidental connection between the two kinds of operators. Thus, when one demands $[a_i,a_i^{\dagger}]=0$, one is not really guaranteeing the Bose-Einstein statistics, but only ensuring that the action of these operators on the energy eigenstates of the two-dimensional isotropic harmonic oscillator is the same in both commutative and noncommutative cases. Hence, we conclude that the relation Eq. (\ref{theta_eta_BE}) cannot accurately describe the relation between noncommutative parameters.

On the other hand, the model discussed in this work does not define any absolute relation between the two noncommutative parameters on the plane. However, by means of the effective Planck constant, it poses a Lorentz invariant constraint on some of the components of the 
noncommutative parameters in 4 dimensions, which is compatible with  the present experimental results.

In summary, we have studied in this work a model where the effect of both configuration and momentum spaces noncommutativity was considered for the two-dimensional gravitational quantum well. The latest results from the experiment by Nesvizhevsky \emph{et al.} allow to bound the fundamental momentum scale introduced by noncommutativity to be below $1\ \mathrm{meV/c}$. Further improvements in the experimental precision could lead to the minimum upper bounds of order $1\ \mathrm{\mu eV/c}$. We find that, to leading order, noncommutativity in configuration space does not affect the energy spectrum of the system. By assuming that the latter introduces a fundamental length scale smaller than the average neutron size, we can conclude that the model modifies the Planck constant by a factor which is at least $24$ orders of magnitude smaller than its value, which is experimentally consistent. The maximum achievable energy resolution in the Nesvizhevsky \emph{et al.} experiment may lower this bound by a factor of about $10^{-5}$. This modification turns out to be Lorentz invariant, and constrains the components of the noncommutative parameters $\theta^{\mu\nu}$ and $\eta^{\mu\nu}$ in a totally different way with respect to the approach proposed in  Ref. \cite{Zhang_1}.

Given that noncommutativity is an attractive theoretical concept which is not fully understood, the studied model introduces some insights on the nature of noncommutativity that are so far consistent with experiments. Possibly, most of the effects of noncommutativity may reveal themselves when our experimental capabilities approach the string scale, which may not happen in the foreseeable future. Meanwhile, low-energy experiments, as the one considered in this work, can help to constrain these effects and hopefully shed some new light on the physical reality.

\vfill

\centerline{\bf {Acknowledgments}}
C.M.L. de A. thanks the Italian Ministero degli Affari Esteri (MAE)
for financial support.
\vskip 0.2cm




\end{document}